# Short echo time Relaxation-Enhanced Magnetic Resonance Spectroscopy reveals broad downfield resonances

Sónia I. Gonçalves, Clémence Ligneul, Noam Shemesh*

*Champalimaud Research, Champalimaud Centre for the Unknown, Lisbon, Portugal*

**Running head:** Short echo-time MRS for downfield spectroscopy

**Conflict of Interest**: None

**Acknowledgements:** The authors thank Prof. Jean-Nicolas Dumez (CEA, France) for his assistance with SLR RF shape algorithm. This study was funded in part by the European Research Council (ERC) under the European Union's Horizon 2020 research and innovation programme (grant agreement No. 679058 - DIRECT-fMRI).

*Corresponding author
Dr. Noam Shemesh, Champalimaud Research, Champalimaud Centre for the Unknown
Av. Brasilia 1400-038, Lisbon, Portugal
E-mail: noam.shemesh@neuro.fchampalimaud.org; Phone number: +351 210 480 000 ext. #4467

Word count: ~3500.

Keywords: Downfield spectroscopy, MRS, Magnetic Resonance Spectroscopy, Exchange, Relaxation Enhancement

**List of abbreviations**

ATP – Adenosine triphosphate
CHESS – Chemical Shift Selective Saturation
iRE-MRS – ISIS-based-Relaxation-Enhanced MRS
ISIS – Image-Selected-In vivo-Spectroscopy
LASER – Localization by Adiabatic Refocusing
MC – Metabolite cycling
MRS – Magnetic resonance spectroscopy
NAA – N-acetylaspartate
NAD – Nicotinamide adenine dinucleotide
NADH – Reduced form of NAD
OV – Outer volume
OVS – Outer volume suppression
PRESS – Point Resolved Spectroscopy
RE-MRS – Relaxation-Enhanced MRS
SNR – Signal-to-noise ratio
SPECIAL – Spin-Echo Full-Intensity Acquired Spectroscopy
STEAM – Stimulated Echo Acquisition Mode
TE – Echo time
TR – Repetition time
VAPOR – Variable power radiofrequency pulses with optimized relaxation delays
VOI – Volume-of-Interest
WATERGATE – Water Suppression by Gradient Tailored Excitation
WEX – Water exchange
WS – Water Suppression

# Abstract

**Purpose:** Most Magnetic resonance spectroscopy (MRS) pulse sequences rely on broadband excitation with water saturation and typically focus on upfield signals. By contrast, the downfield spectrum, which contains many potentially useful resonances, is typically not targeted since conventional water-suppressed techniques indirectly saturate the labile protons through exchange. Relaxation-Enhanced MRS (RE-MRS) employs frequency-selective excitation while actively avoiding bulk water perturbation, thereby enabling high-quality downfield spectroscopy. However, RE-MRS typically requires very long (typically >40 ms) echo times (TEs) due to its localization module, which inevitably decreases sensitivity and filters shorter $T_2$ components. Here, we overcome this limitation by combining RE-MRS and Image Selected In vivo Spectroscopy (ISIS) localization, abbreviated iRE-MRS, which in turn allows very short TEs (5 ms using our hardware).

**Methods:** Experiments were performed in vitro for validation as well as and in *in vivo* brain at 9.4 T.

**Results:** The new iRE-MRS methodology was validated in phantoms where good performance was noted. When the downfield spectrum was investigated at short TEs in *in vivo* rat brains, iRE-MRS provided very high sensitivity; the ensuing downfield spectra encompassed numerous broad peaks, as well as a broad baseline. All downfield spectral peaks were highly attenuated by increasing TEs as well as by applying water saturation, though to different extent. The signal ratios also varied between TEs, suggesting that exchange rates are different among the downfield signals.

**Conclusions:** Short-TE iRE [1]H downfield MRS opens new directions in the investigation of *in vivo* downfield metabolites and their role on healthy and disease processes.

# Introduction

Magnetic resonance spectroscopy (MRS) is a versatile technique reflecting endogenous metabolite levels *in vivo*. Since MRS can detect signals originating from small molecules (typically at a ~mM concentration) that are involved in metabolism and/or are localized within particular cell types or subcellular spaces, it can potentially provide enhanced specificity, thereby informing on biochemical and physiological processes related to tissue function and mechanisms of disease (1, 2). Most MRS pulse sequences localize volumes of interest using a combination of broadband, slice-selective pulses. For example, the prevalent Point Resolved Spectroscopy (PRESS) (3, 4), Stimulated Echo Acquisition Mode (STEAM) (5) or Spin-Echo Full-Intensity Acquired Spectroscopy (SPECIAL) (6) all rely on broadband pulses with slice selection to define the voxel. Due to the broadband nature of the RF pulses, a water-suppression (WS) module such as CHESS (7, 8) or VAPOR (9) is required prior to magnetization excitation to acquire "clean" spectra. By contrast, methods such as Localization by Adiabatic Refocusing (LASER) (10) and Image Selected In Vivo Spectroscopy (ISIS) (11) can avoid broadband slice-selective excitation pulses: LASER achieves the volume selection with three pairs of adiabatic refocusing pulses following spectrally-selective excitation; though benefiting from an improved localization profile (12), LASER cannot achieve very short echo times due to the long duration of typical adiabatic pulses. ISIS, on the other hand, localizes the Volume-of-Interest (VOI) using slice-selective inversion pulses that follow a Hadamard scheme (13) thereby canceling signals outside of the VOI (14), followed by a spectrally-selective broadband excitation. Typically, the short excitation pulse will be immediately followed by acquisition, furnishing ISIS with very short TE capabilities and making it particularly useful for acquiring signals arising from short $T_2$ spins, such as e.g., $^{31}P$ (15-19).

The vast majority of MRS studies explored the spectral region upfield of water, which contains, among others, nonoxygenated aliphatic protons, aliphatic/aromatic acetoxyl protons, as well as methoxy protons (20). However, the spectrum downfield of the water signal is very rarely investigated by MRS, even though it contains highly relevant chemical functional groups, such as hydroxyls, amines, amides, aldehydes, aromatic, and carboxylic protons, among others (21). Since some of these biologically-relevant functional groups are present at relatively high concentration *in vivo*, they could provide complementary information to the upfield spectrum. For example, phenylalanine, homocarnosine, adenosine triphosphate (ATP), NADH, or glutathione all generate signals in the downfield spectrum (22). Therefore, the development of downfield MRS methods, as well as the characterization of the ensuing signals in terms of concentration and relaxation/exchange rates, and peak assignments, can be important for a more complete understanding of tissue metabolism.

Perhaps the main reason for the scarcity of downfield MRS studies is the widespread application of WS modules, which are a-priori deleterious for exchangeable resonances due to direct exchange of saturated water protons with the downfield resonances of interest. MRS without active WS has been proposed in the past and achieved using broadband excitation, high dynamic range MRS. Water Suppression by Gradient Tailored Excitation (WATERGATE) (23) employs a combination of gradients and water-selective RF pulses to selectively dephase bulk water signals while retaining a full spin-echo from metabolites. The study in (24) used WATERGATE in rodents and showed that, by avoiding water pre-saturation, several pH-sensitive downfield resonances could be detected. van Zijl et al. (25) applied a water exchange (WEX) filter to study downfield proton exchange dynamics both *in vitro* and *in vivo*. Gradient cycling (26) is a two-scan acquisition that offers water-free spectra corrected for side band artifacts (27). Finally, Metabolite Cycling (MC) (28) is another two-scan acquisition that has been applied in two of the first human studies focusing on downfield resonances in human brain and skeletal muscle, respectively (29, 30). More recently, human studies have been extended to 7 and 9.4 T using either a combination of water suppression (31) or MC (32) with short TE STEAM to measure the $T_1$ and $T_2$ values as well as concentration of downfield peaks. Some of the downfield signals have been tentatively assigned, in particular, the amide peak around 7.8 ppm was assigned to NAA (though other amides cannot be a-priori ruled out, homocarnosine at 7.05 and 8.02 ppm, α-glucose at 5.22 ppm and NAD$^+$ in the range 8.7-9.5 ppm (22, 31, 33).

Relaxation-Enhanced MRS (RE-MRS) (34, 35) has been recently proposed as a powerful method for downfield spectroscopy. RE-MRS avoids water pre-saturation by foregoing broadband perturbations, rather replacing them with frequency-selective pulses targeting only resonances (or spectral bands) of interest. The bulk water is, by RF pulse design, left nearly unperturbed and therefore does not attenuate signal arising from exchanging spins. Furthermore, in some cases, the large reservoir of unperturbed water can induce "Relaxation Enhancement": polarization transfer and/or exchange that contributes to an apparent decrease in metabolite $T_1$s. Such RE-MRS experiments have been used for studying the downfield spectrum of fresh *ex-vivo* rat brain and revealed relaxation enhancement effects (34). RE-MRS was then extended to *in vivo* settings, where localization has been achieved using LASER, thereby enabling characterization of the downfield spectrum as well as providing a useful platform for imparting different contrast mechanisms, such as Double-Diffusion-Encoding (36), for metabolic signals (35). de Graaf and Behar (33) harnessed a similar RE-MRS approach to detect NADH and NAD signals downfield and confirmed the first relaxation enhancement observations (34).

Downfield resonances are typically characterized by relatively short transverse relaxation times, mainly due to water exchange-mediated line broadening. Short TE acquisition schemes could thus potentially "rescue" TE-driven attenuation and increase the signal-to-noise ratio (SNR) of the measured signal. However, LASER localization is typically very demanding in terms of TE: RE-MRS at 21.1T

(34) was performed using TEs around 60 ms. de Graaf and Behar (33) managed to reduce the TE at 11.7 T to 14 ms by employing minimum-phase excitation pulses (37) and very short adiabatic refocusing pulses made possible using specialized hardware (33). Given that many of the downfield linewidths can approach (at least) ~50-100 Hz at 9.4 T (32), there would be a clear incentive for achieving even shorter TEs for downfield spectroscopy.

Here, we introduce a new method for [1]H downfield spectroscopy that combines RE-MRS principles and ISIS localization, abbreviated iRE-MRS, which, when combined with carefully optimized outer volume suppression (OVS), provides high sensitivity *in vivo* downfield spectra at 9.4 T with TEs as low as 5 ms. Following validation in phantoms we show how WS affects short TE downfield signals as well as explicit TE dependence in the rat brain *in vivo*. Potential future applications are discussed.

# Methods

All experiments involving animal care strictly followed experimental procedures in agreement with Directive 2010/63 of the European Parliament and of the Council, were preapproved by the Champalimaud Animal Welfare Body and the national competent authority (Direcção Geral de Alimentação e Veterinária, DGAV).

**Pulse Sequence.** The proposed iRE-MRS pulse sequence is shown in Figure 1 for one ISIS cycle. It is based on spectrally-selective excitation and refocusing, the latter typically required to refocus chemical shift evolution during the excitation pulse. The refocusing pulse is surrounded by weak spoiler gradients along the three physical axes to suppress unwanted coherence pathways and residual water signals. Localization is achieved by concatenating an ISIS block, consisting of a combination of three adiabatic spatially-selective inversion pulses, prior to the spin-echo sequence (the refocusing is necessary due to chemical shift evolution in our linear-phase excitation pulses). In addition, OVS modules were also implemented to improve the localization quality.

**RF pulse design.** RF pulses were designed with the SLR algorithm (38) using software implemented in Matlab (Mathworks, Natick, Mass). In some experiments (see below), the RF pulses were generated by the shape algorithm implemented in Paravision 6.0.1 (Bruker Biospin, Ettlingen, Germany), which also utilizes an implementation of the SLR algorithm. Pulse-specific details are provided below for each experiment.

### MRS experiments

All experiments were performed on a 9.4T horizontal bore scanner (Bruker Biospin, Karlsruhe, Germany) equipped with an Aeon III HD console and with gradients capable of producing up to 660 mT/m isotropically. An 86 mm quadrature resonator was used for excitation while a 4-element array cryocoil (Bruker BioSpin, Fallanden, Switzerland) was used for signal reception.

**In vitro experiments.** To validate the iRE-MRS pulse sequence, experiments were performed on a phantom containing two tubes immersed in Fluorinert (Sigma Aldrich, Lisbon, Portugal), filled with 100 mM solutions (in Phosphate Buffered Saline (PBS), pH = 7.5) of NAA and GABA respectively, at room temperature.

iRE-MRS spectra were acquired in $2.2^3$ mm$^3$ voxels localized within each tube. The excitation (8 ms equiripple single band) and refocusing (5 ms equiripple) pulses were centered at 2 ppm, with a 3 ppm bandwidth (Figure 2a), to include the resonance frequencies of NAA and GABA resonating in this spectral region. Both RF pulses were generated with the SLR algorithm implemented in Matlab.

iRE-MRS data were acquired using the following sequence parameters: TR/TE = 40000/15.3 ms, acquisition bandwidth = 4807.69 Hz, 7211 data points, spoiler gradient of 300 μs and 13.33% of the maximum gradient amplitude ($G_{max}$ = 660 mT/m), 2 repetitions (each repetition consisting of 8 ISIS steps), the total scan duration was 10 mins 40 s. In addition, "gold standard" spectra were acquired with PRESS, attempting to use as many equivalent parameters as possible. With the exception of the amplitude and duration of spoiler gradients, all other acquisition parameters were identical to iRE-MRS. VAPOR WS was used for the PRESS acquisitions, and the timings of the VAPOR pulses and delays were manually optimized to reduce residual water signal.

***In vivo experiments.*** *In vivo* experiments were carried out on Long-Evans female rats (N = 8) weighting approximately 200 g and aged between 10 and 12 weeks. The animals were housed in a 12 h light/dark cycle, two per cage. Anesthesia was induced using 5% isoflurane in 100% $O_2$ and kept at 1.5-2% isoflurane for the duration of the experiment. Temperature and breathing rate were constantly monitored using PC-SAM software (SA Instruments, NY, US).

**Upfield experiments.** iRE-MRS spectra were obtained in a centrally-positioned 6.0×2.5×3.5 mm$^3$ voxel, predominately including cortex and caudate-putamen. The RF excitation (Figure 2a) and refocusing pulses were designed by a Matlab implementation of the SLR algorithm (38) to have a bandwidth of 3 ppm (with a 1% ripple on the water signal at 4.7 ppm). Both pulses were centered at 2 ppm and their transition band was approximately equal to 0.9 ppm. TR/TE = 15000/16 ms and 4 repetitions, each containing 8 ISIS steps, were acquired, resulting in a total scan duration of 8 mins. The spoiler gradient was set to 300 μs and 13.33% of the maximum gradient amplitude; the spectral width was 4807 Hz, 1442 data points. To compare the iRE-MRS data with the gold standard, reference spectra were acquired in the same voxel using PRESS, VAPOR water suppression (manually optimized), OVS module, and identical TEs, TRs, number of repetitions and spectral acquisition parameters. In order to make it comparable, the number of repetitions in PRESS was set to 8 times the number of repetitions in iRE-MRS.

**Downfield experiments.** iRE-MRS data of the downfield part of the spectrum were obtained in the same voxel as for upfield experiments. The pulses used in these experiments were generated by the SLR algorithm implemented in Paravision 6.01 (Bruker Biospin, Ettlingen, Germany), as these RF shapes allowed for a shorter TE at the cost of a slightly broader RF profile. Both excitation and refocusing pulses were centered at 9.5 ppm, and had a bandwidth of 6.5 ppm (Figure 2b), with a residual water excitation of less than 6%. TR/TE = 15000/5 ms, and 15 repetitions were acquired (each repetition containing 8 ISIS cycles), resulting in a total scan duration of 30 mins. The spoiler gradients were set to 300 μs and 13.33% of the maximum gradient amplitude and the spectral width for these acquisitions was 5597.01 Hz, 1442 data points. Additionally, high SNR downfield spectra were acquired for N = 3 rats, each spectrum acquired with 30 repetitions, resulting in a total scan duration of 1 hour. All other acquisition parameters remained as described above.

**Experiments varying TE.** The effects of echo time were investigated by acquiring two additional experiments with longer TEs (16 and 40 ms) for N = 3 rats. All other acquisition parameters remained as described above.

**Effects of Water Suppression.** The effects of WS pulses on downfield spectra was investigated in N = 2 rats. iRE-MRS data were acquired at the minimum TE=5 ms with and without WS RF pulses. When WS was applied, a VAPOR pulse sequence was included prior to iRE-MRS's excitation, and, similarly to PRESS, the timings were manually adjusted to minimize residual water signal.

**Data processing.** Post-processing was reduced to a minimum: each FID (resulting from averaging over 8 ISIS steps, 1 repetition) was Fourier transformed, individually rephased and averaged over the total number of repetitions acquired for each scan. For inter-individual comparison, spectra were scaled with respect to the water peak intensity in unsuppressed spectra. Apodization was applied only when explicitly indicated. To create a representative iRE-MRS downfield spectrum, data from N=3 rats were first aligned, taking the "NAA" peak as reference, z-scored and then averaged.

# Results

Validation experiments in the phantom are shown in Figure 3. iRE-MRS successfully spatially resolved GABA (Figure 3a) and NAA (Figure 3b), with no observable contamination – whether spatially from the other metabolite tube or spectrally from water. The spectral quality in iRE-MRS experiments was comparable to the gold-standard PRESS experiments within the targeted bandwidths. Multiplets corresponding to $^2CH_2$, $^3CH_2$ and $^4CH_2$ GABA resonances (Fig. 3c) as well as NAA's prominent methyl singlet (Fig. 3d) were well-reproduced (19) using iRE-MRS. Phase differences between PRESS and iRE spectra, attributed to the different J-evolution in each sequence, were noted for these multiplets. The localization performance and spectral fidelity of iRE-MRS was thus deemed comparable to PRESS.

*In vivo* iRE-MRS data are shown in Figure 4 for a representative rat (Rat 5). The voxel position is shown in Figure 4a. To assess iRE-MRS's capability in an *in vivo* setting, upfield experiments were performed and compared to the gold-standard PRESS (Figure 4b). Within the pulse passband (0.5-3.5 ppm), the iRE-MRS spectrum agrees well with its PRESS counterpart, while the peaks that are located towards the edge of the passband are attenuated by the transition band of the RF pulse. The signal outside the excitation range is dominated by noise, which experimentally confirms the sharpness of the RF excitation profile and effectiveness of the crushers. Very little, if any, water contamination was noted.

Representative iRE-MRS downfield experiments at TE = 5 ms are shown in Figure 4c. The spectral quality was very good, and the signal-to-noise ratio in this animal was 20.6, of the same order of magnitude of the SNR in PRESS (20.7) and iRE-MRS upfield (27.1).

Figure 5 shows the TE dependence of downfield iRE-MRS spectra for 3 different animals. The signal decay with increasing TE was quite strong, with peaks *a*, *b*, *c*, *d* and *e*, attenuating by approximately 66%, from TE = 5 to TE = 16 ms; these peaks were nearly undetectable at TE = 40 ms.

Note also that the ratios between the signals and, for example, the amide peak at 7.8 ppm vary considerably with TE. In addition, it is worth noting that iRE-MRS spectra showed good reproducibility over different animals for the various TEs used here (Figure 5).

To better characterize the short TE downfield spectral resonances, an additional experiment targeting higher SNR was performed (Figure 6). Figures 6a, 6b and 6c show the individual spectra obtained in three animals; good reproducibility was apparent, and the signal to noise ratio of the 'NAA' peak was 57.6±9.7. Figure 6d shows the average spectrum over three animals. The peaks labeled as "*NAA*", *a*, *b*, *c*, *d*, and *e* were easily distinguishable, while other smaller peaks became more apparent in this spectrum, e.g., at ~8.5 (*f*), 8.25 (*g*), ~9.0 (*j*), 9.3 (*i*) and 9.5 (*h*) ppm. No other spectral peaks were detected beyond 10 ppm.

The effect of water saturation on short-TE downfield spectra is presented in Figure 7 for one representative rat. As expected, the signal intensity of spectral peaks *a-e* is strongly attenuated (for peaks *a* and *b* by more than 80%) when water suppression was applied. The signal amplitude of the "NAA" peak at ~7.8 ppm, with respect to the baseline signal, remains approximately unchanged in both datasets, indicating slower exchange rate with water protons.

To assess whether iRE MRS could be used for dynamic experiments, Figure 8 shows a downfield iRE-MRS spectrum, obtained in one animal for TE = 5 ms, acquired in just 2 minutes (8 ISIS cycles). The SNR for this acquisition was 34.7 for the amide 'NAA' peak at 7.8 ppm. Most spectral signatures that were identified in higher SNR spectra could still be detected using this rather rapid acquisition.

# Discussion

Methods for reducing echo time in [1]H MRS have been vigorously developed over the years, resulting in STEAM, SPECIAL and (mainly for [31]P) ISIS pulse sequences (5, 6, 11). The exchanging nature of downfield signals and the inherent line broadening associated with exchange suggests that downfield MRS can significantly benefit from TE reduction. In previous studies, short-TE MRS (40) has already been used for downfield MRS with minimal TEs of the order of 10-13 ms (30, 31). However, broadband excitation and STEAM were necessary, carrying a 50% signal penalty and a mixing time period that could facilitate proton exchange with saturated water. de Graaf and Behar (33) reached TEs as short as 14 ms using minimum phase pulses and LASER localization but harnessing unconventional high-performance gradient systems and RF power.

Here, we presented iRE-MRS: a combination of RE-MRS – a method obviating the water suppression and its associated deleterious effects on downfield signals – and ISIS – a method achieving localization with short TE. The RE-MRS component includes frequency-selective excitation which affords a better dynamic range and hence higher sensitivity for the detected signal. On the other hand, ISIS's use of longitudinal magnetization for spatial localization provides access to short echo times that preserves signals originating from shorter $T_2$ species and enhancing signal to noise ratios. Thus, iRE-MRS is capable of providing high-sensitivity and short TE downfield spectra.

The *in vitro* study in this work was designed to validate iRE-MRS (Figure 3). Indeed, iRE-MRS spectra reproduced the gold-standard PRESS spectra with high accuracy. The slight phase difference in the multiplets between iRE-MRS and PRESS spectra (e.g. Figure 3d) are expected (34, 35, 39) and can be traced to different J-coupling evolutions in the two pulse sequences (35) that originate from their different refocusing schemes (data not shown). The ensuing *in vivo* iRE-MRS validations also showed high quality upfield spectra which were comparable between PRESS and iRE-MRS (Figure 4). The high a-priori sensitivity is also in part due to our use of a 4-element array cryocoil, which enhances the signal relative to a similar room-temperature coil by ~2.5 at 9.4T (41). The shortest achievable TE in our current experimental setup was 5 ms; incidentally, this is the shortest TE we are aware of that has been insofar applied for downfield MRS.

Following these validations, iRE-MRS was used to investigate the *in vivo* downfield signals in healthy rat brains at 9.4 T. The spectral quality of iRE-MRS was good, showed high reproducibility (Figure 5) and iRE-MRS could be even obtained within ~2 min from a $6.0 \times 2.5 \times 3.5$ mm$^3$ voxel *in vivo* (Figure 8). Importantly, in addition to the "NAA" amide peak at ~7.8 ppm (placed in quotation marks since other amides may also reside on the same resonance), multiple other peaks could be observed at ~6.8, 7.1. 7.3, 7.5 and 8.3 ppm (Figure 5), which largely confirm results previously obtained in both animals and humans (29-35); however, the typically observed ratios of some of these peaks was very different at short TE when compared to longer TEs used in other studies (29-35). Indeed, when the TE was prolonged from 5 ms to 16 ms, peak ratios already appear very different (Figure 5). Similarly, actively suppressing the water signal results in varying degrees of attenuation (Figure 7) with the peak 'a' undergoing the largest effect. All these are suggestive of varying degrees of exchange in these downfield signals, and underscore the importance of short TE for downfield MRS.

The spectral peaks appearing between 6.5 and 7.5 ppm ('b', 'c', 'd' and 'e' in Figure 6) probably originate from metabolites such as ATP, creatine, phosphocreatine or glutamine (31). In the spectral range from 7.5 to 9 ppm, the "NAA" amide peak was detected in all animals. The multiple peaks in the range between 8 and 8.5 ppm (labeled 'a', 'f' and 'g' in Figure 6) likely arise from amide protons in peptides/proteins and/or metabolites such as glutathione and homocarnosine (42). The spectral range from ~8.8 to 10 ppm exhibits three small peaks (labeled 'h', 'i' and 'j' in Figure 6) that probably correspond to the H2, H6 and H4 proton resonances in NAD$^+$ previously reported at a higher field strength (33). However, the broad signal of peak *a* at TE = 5 ms partly masks these signals, implying that other resonances probably underlie these spectral regions. Further studies are required to elucidate the molecular origins of the signals at different TEs.

Although ISIS can produce very short TE spectra that are relatively independent of J-coupling and relaxation effects, it has very seldom been used in [1]H MRS. This is likely due to the fact that ISIS localization requires 8 individual scans, which makes it highly sensitive to OV contamination, in case of incomplete signal cancellation. One of the main sources of OV contamination is motion during the 8 scans, and, for this reason, good animal immobilization is crucial. OV contamination can also occur if magnetization is not fully relaxed, a situation in which the add-subtraction scheme becomes dependent on the acquisition order and on the ratio $TR/T_1$ (11, 43, 44). This problem is more severe for long $T_1$ metabolites, but it can be circumvented by using sufficiently long TRs, taking as reference the longest $T_1$ in the specimen, or mitigated by choosing a scan order that minimizes OV contamination (12). OV contamination can also result from the transition bands of imperfect ISIS inversion pulses and/or from imperfect excitation pulses. While the use of adiabatic inversion pulses minimizes the first issue, imperfect excitation RF profiles are of concern, especially for shorter TRs where residual transverse magnetization remains when the following scan begins. This problem would in principle be eliminated if long TRs are used. In this case, even if the excitation profile does not yield a spatially homogeneous 90° flip angle, the signal outside the VOI will still be cancelled if the $B_1$ map remains stationary in time. Indeed, OVS optimization was critical for this study. Multidimensional RF pulses (45-48) that simultaneously encode spectral and spatial dimensions during both excitation and refocusing could be used in the future to further mitigate such issues since they could be used to replace the ISIS encoding in two spatial dimensions.

It is interesting to consider exchange effects *during* the ISIS module in iRE MRS. Clearly, selective inversion of water with an odd or even number of pulses could potentially propagate to the downfield signals via inversion transfer. However, since the ISIS module employs broadband, nonselective inversion, the downfield signals are inverted the same number of times as the water signal. Barring imperfections, if the $T_1$s were equal, this mechanism would not incur signal loss. On the other hand, if the $T_1$ of one of the moieties is much shorter than the other and the ISIS delays are long compared to $T_1$, then signal loss may occur due to exchange. The $T_1$ values of several downfield metabolites are in the range 400-700 ms (31), shorter than the water $T_1$ at ~2000 ms at 9.4 T (49). However, it has been shown previously (33, 34) that, non-selective magnetization inversion increases downfield metabolite $T_1$s by a factor of 3-4 due to magnetization exchange with water. In our experiments, the duration of the entire ISIS module was approximately equal to 20 ms, much shorter than the average metabolic and water $T_1$s even given selective inversion. Hence, this mechanism of signal loss is likely not a major confounding factor of this study, though it should be considered if ISIS durations become much longer.

In the past, RE-MRS was proposed to enhance the signal-to-noise per unit time by facilitating relaxation enhancement which, in turn, enabled very large TR reductions. iRE-MRS cannot efficiently exploit such TR reductions since the ISIS component then becomes vulnerable to OV contamination as mentioned above. However, it seems that iRE-MRS's TE reduction is well worth the associated long TRs, given the excellent sensitivity it produces.

It is worth considering further improvements to iRE-MRS. For example, minimum phase excitation pulses (37) could be used to avoid the refocusing pulse, thereby decreasing the echo time significantly. At higher magnetic field strengths, iRE-MRS will profit from the inherent larger spectral separation and even higher SNR, but transverse relaxation rates will decrease. The optimum magnetic field for these tradeoffs remains to be investigated. It should be noted that higher magnetic fields also require stronger RF amplifiers to achieve the same flip angles. Future work should also target a more detailed characterization of downfield metabolites: e.g. spectral identification, concentration calculation, relaxation and exchange rates measurement. Finally, the use of short-TE iRE-MRS at higher field strengths, benefiting from the inherent increase in SNR and spectral separation, has the potential to unravel new directions in the investigation of *in vivo* downfield resonances, especially vis-à-vis their potential to characterize metabolism in both healthy and diseased states.

# Conclusions

The applicability of [1]H ISIS localization for short-TE downfield RE-MRS (iRE-MRS) has been demonstrated for the first time. At the minimum achieved TE in this study of 5 ms, *in vivo* iRE-MRS yields robust downfield spectra with high sensitivity at 9.4 T. Additionally, the inherent sensitivity of iRE-MRS yielded enhanced downfield spectra in as little as 2 minutes, which could potentially provide new interesting ways to dynamically study brain metabolism. The TE dependence and water saturation effects point out to varying degrees of exchange in these signals, which can be interesting for (a) more accurately quantifying the downfield spectra and (b) may have applications in measuring exchange properties in normal and diseased tissues. All these bode well for iRE-MRS as a new tool for *in vivo* downfield MRS.

# Figures

**iRE-MRS pulse sequence**

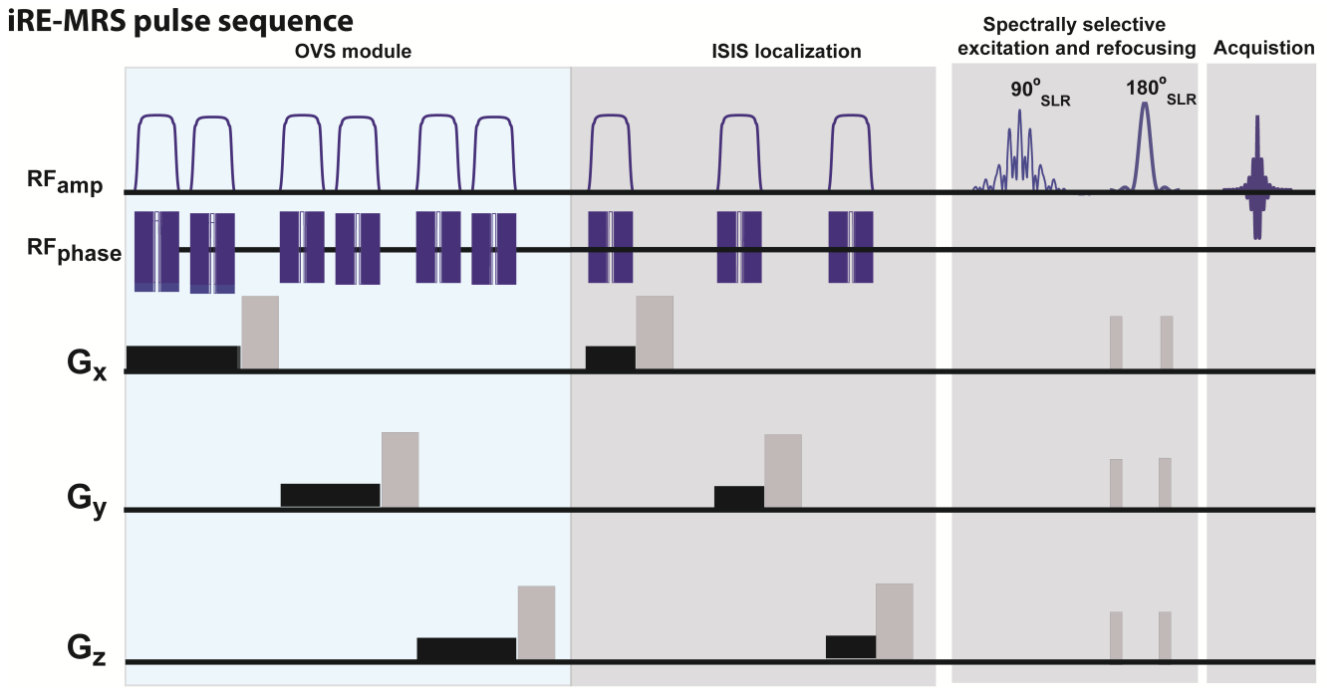

**Figure 1. iRE-MRS Pulse sequence diagram.** The ISIS localization block precedes the RE-MRS sequence, namely, the spectrally-selective excitation and refocusing pulses followed by data acquisition. Note that *in vivo*, outer volume contamination becomes a significant issue due to the ISIS block, thereby necessitating a well-optimized OVS module.

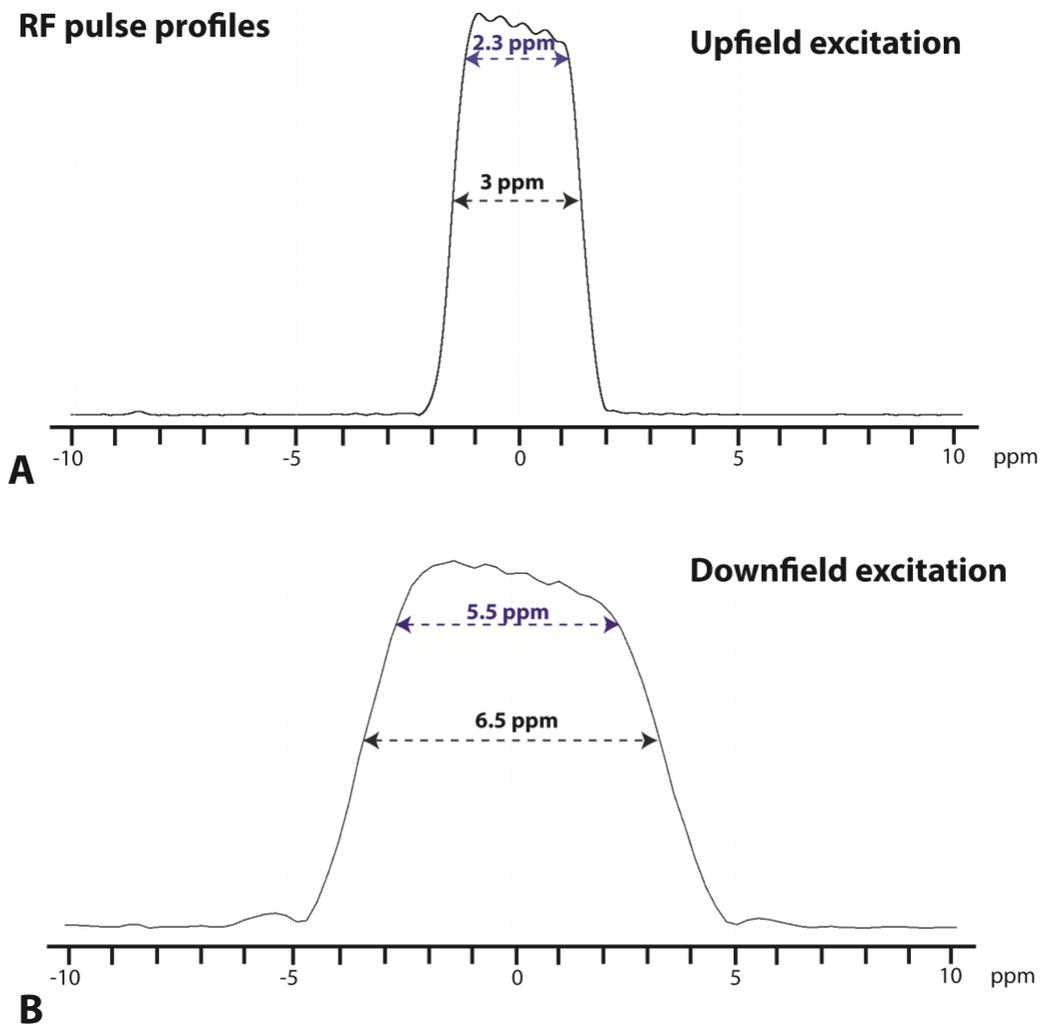

**Figure 2. RF Excitation profiles.** Measured profiles of the RF pulses that were used for upfield (A) and downfield (B) excitation. The FWHM and 90% passbands are indicated with black and blue arrows respectively.

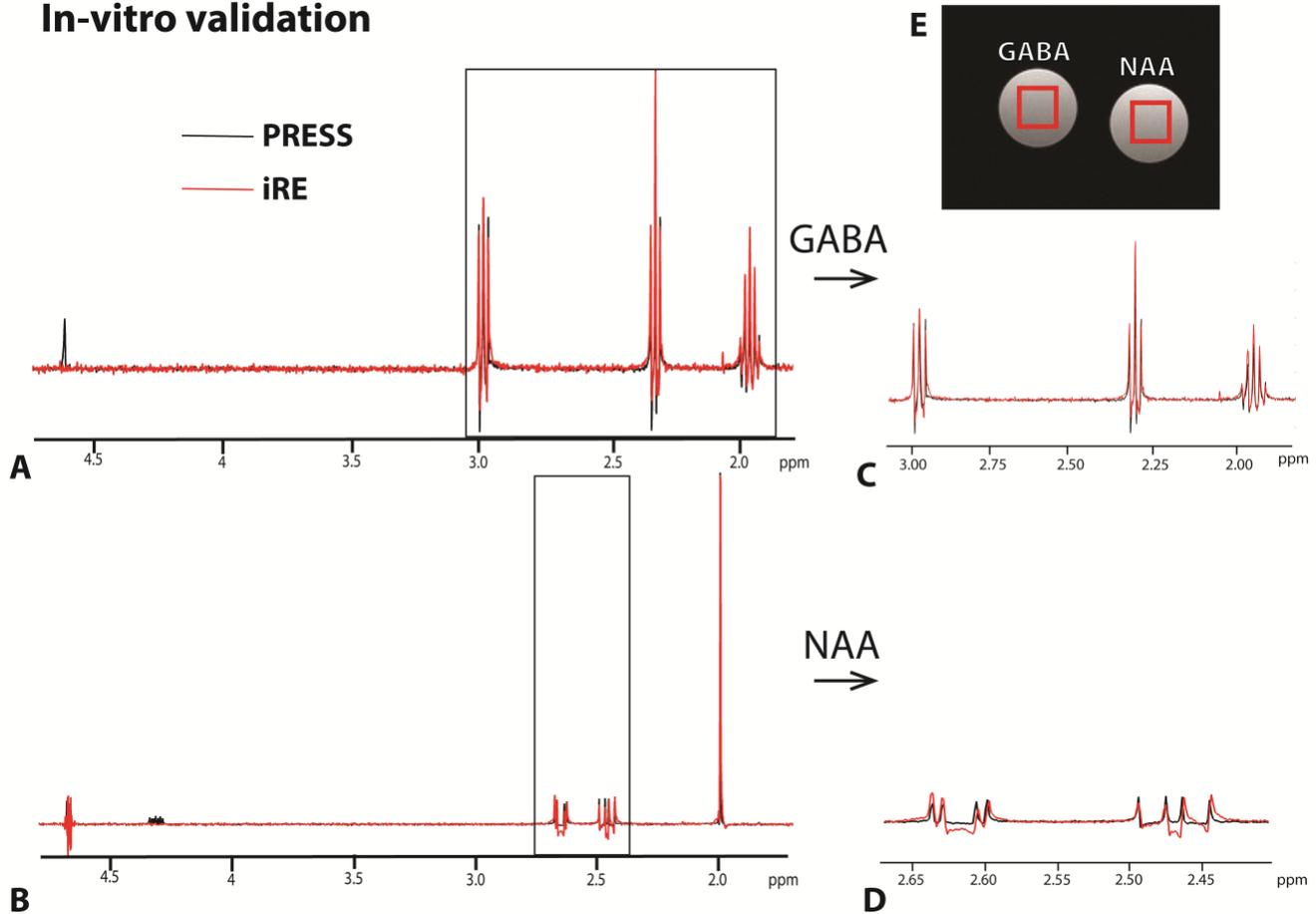

**Figure 3. Phantom experiments.** Spectra obtained with PRESS (black) and iRE-MRS (red) from each of the two tubes containing NAA or GABA, respectively. Acquisition BW=4807.69Hz, Number data points=7211 data points, 2 repetitions, remaining acquisition parameters as described in main text **(A)** Spectrum from the GABA sample. **(B)** Spectrum from the NAA sample. **(C)** Magnification of the shaded area in (A) (2.3-fold magnification). **(D)** Magnification of the shaded area in (B) (4-fold magnification). Note that the iRE-MRS spectra well reproduce the PRESS signals. **(E)** Axial T2-weignted image of the phantom with the representation of the voxel (red square) position.

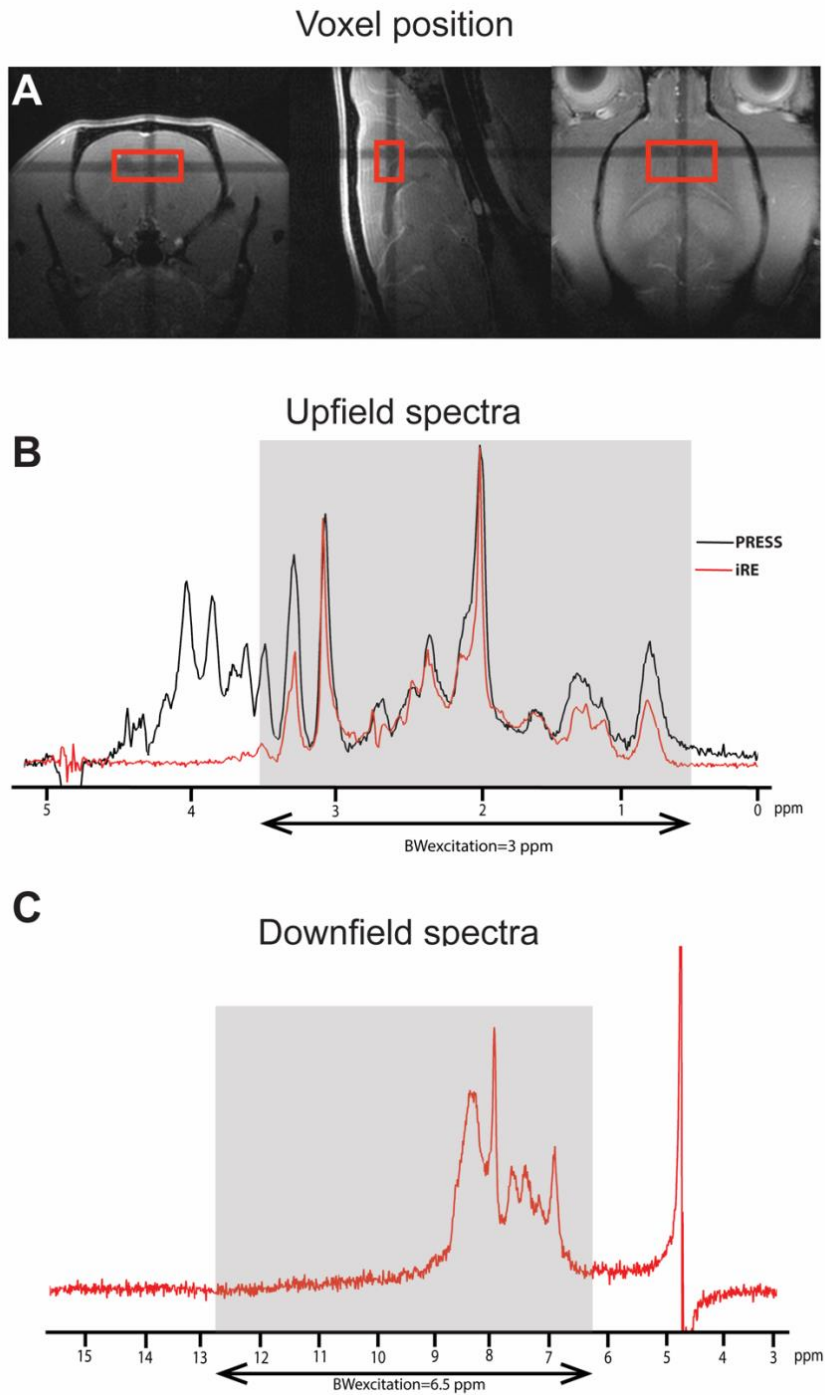

**Figure 4**. *In vivo* **iRE-MRS at short TE. (A)** Voxel position is shown by the red box. **(B)** Representative upfield spectra obtained with iRE-MRS (red trace) and PRESS (black trace) from a representative rat. Acquisition BW = 4807.69Hz, Number data points = 1442, 4 repetitions, remaining acquisition parameters as described in main text. As with the in vitro experiments, the *in vivo* iRE-MRS

spectra reproduce the PRESS spectrum within the spectrally-selective pulse profile. **(C)** Representative downfield spectrum using iRE-MRS and the following parameters: TE = 5 ms, excitation centered at 9.5 ppm, BW = 6.5 ppm, acquisition BW = 5597.01 Hz, Number data points = 1442, 15 repetitions. Remaining acquisition parameters as described in main text.

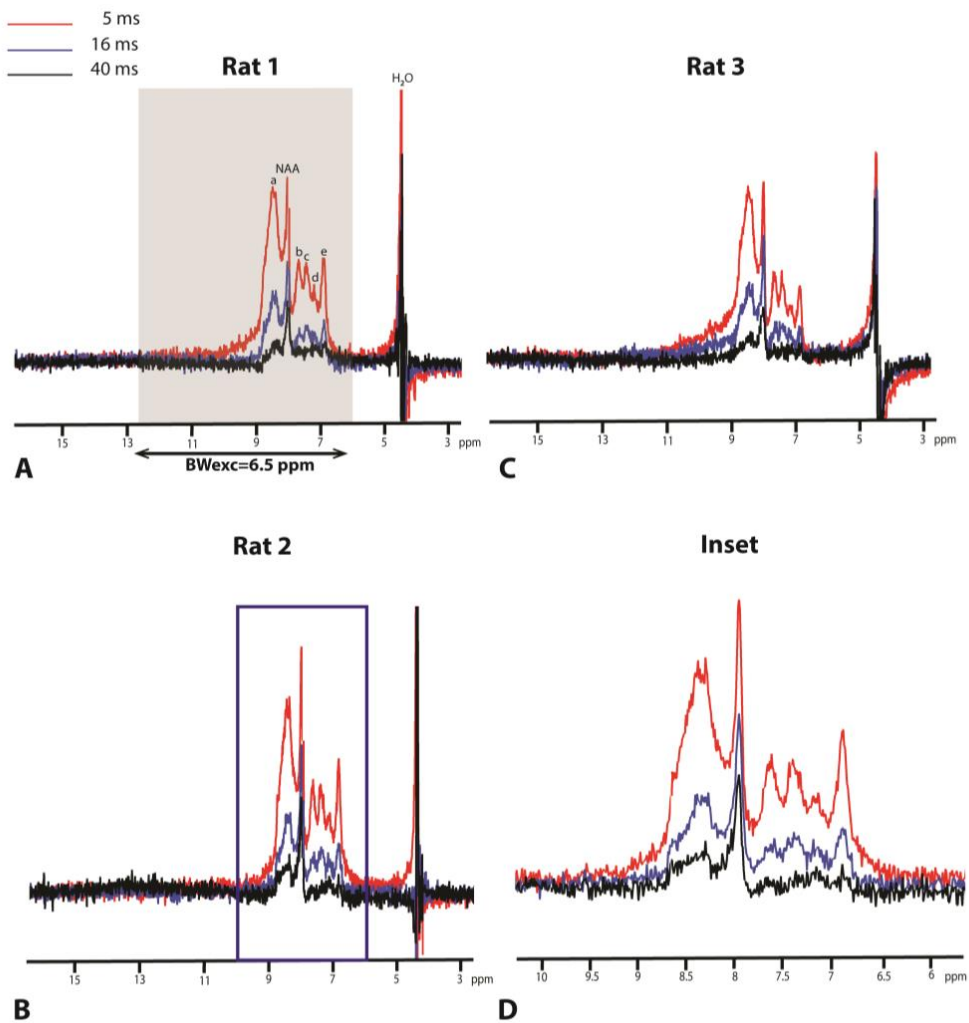

**Figure 5**. **TE-dependence of downfield resonances. (A-C)** Spectra obtained from Rats 1-3, respectively (excitation centered at 9.5 ppm, BW = 6.5 ppm, acquisition BW = 5597.01 Hz, Number data points = 1442, 15 repetitions. Remaining acquisition parameters as described in main text.). Notice the strong attenuation between TE = 5 ms and TE = 16 ms, demonstrating the short apparent $T_2$ of most downfield signals, due to exchange with bulk water. Reproducibility among the three different rats was very good. The spectra for the three animals were acquired with the following SNRs (over the range (6.75-9 ppm): for TE = 5 ms SNR = 16.2, 15.8, 15.0; for TE = 16 ms SNR = 5.4, 6.4, 6.6; for TE = 40 ms SNR = 2.5, 2.0, 1.2. (**D**) Enhanced view of (B).

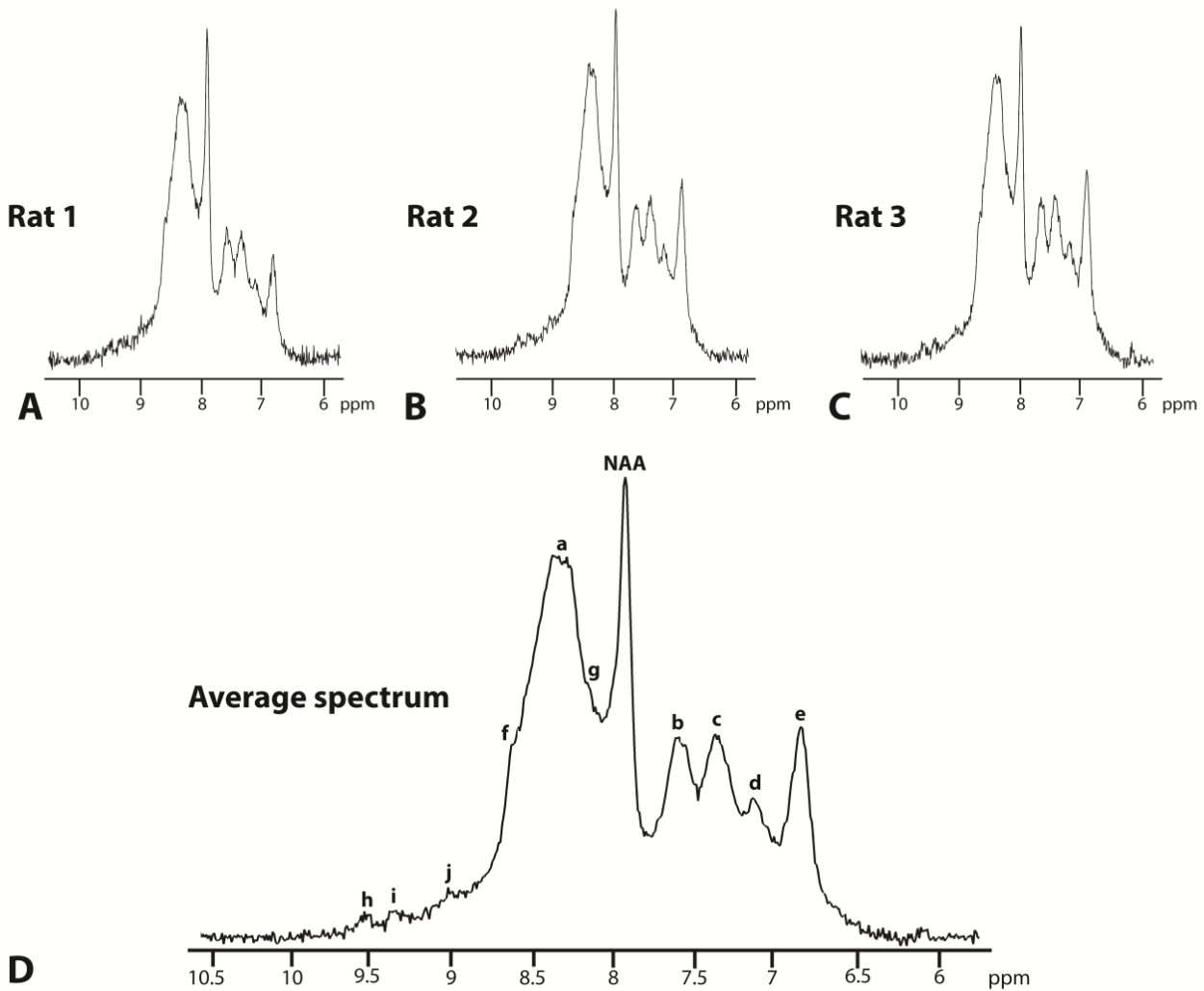

**Figure 6. High SNR experiments.** iRE-MRS downfield spectra from 3 rats (A-C, respectively) using 30 repetitions and the minimum TE of 5 ms (acquisition BW=5597.01 Hz, Number data points=1442, remaining acquisition parameters as described in main text). D) Average spectrum.

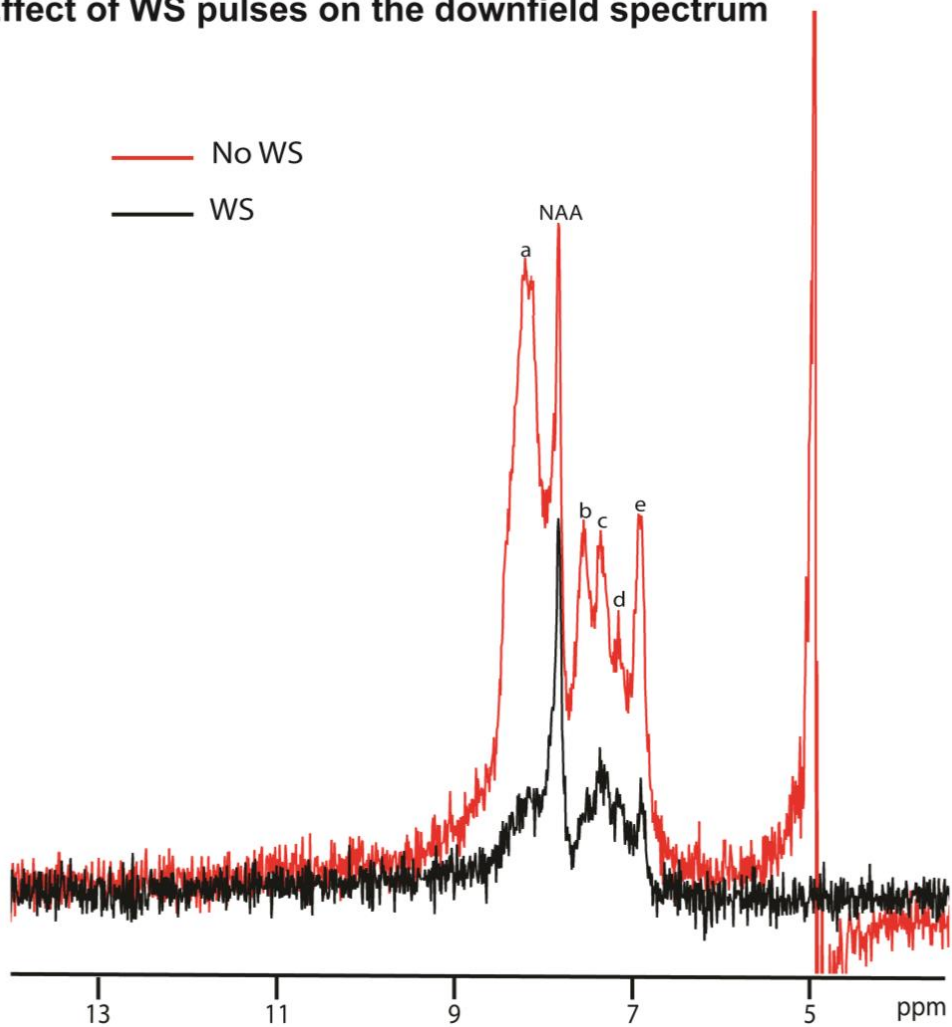

**Figure 7**. **Water saturation effects.** Representative iRE-MRS downfield spectra from a representative

rat showing the effect of water saturation. Very strong attenuation of the downfield signals was apparent at the short echo times (data acquired with 15 repetitions, TE=5 ms, acquisition BW = 5597.01 Hz,

Number data points = 1442, remaining acquisition parameters as described in main text).

**Fast acquisition of a downfield spectrum with iRE-MRS**

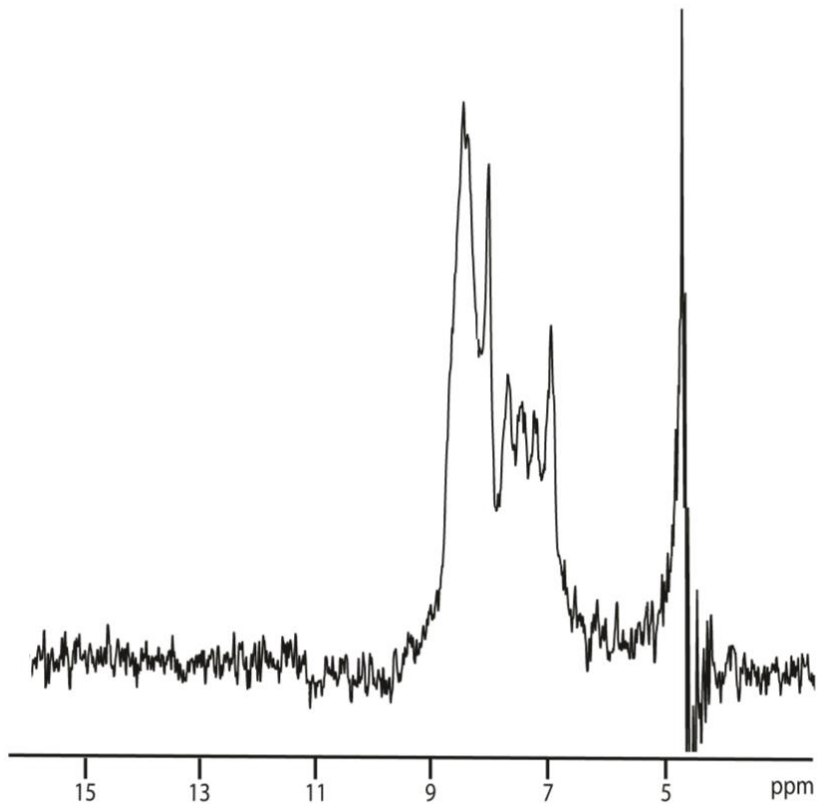

**Figure 8. Rapid acquisition of downfield spectra.** iRE-MRS downfield spectrum from a representative rat at TE = 5 ms, 1 repetition, acquisition BW = 5597.01 Hz, Number data points = 1442, 20 Hz apodization, remaining acquisition parameters as described in main text. Total scan time = 2 mins.